\documentclass[11pt,a4paper]{article}

\usepackage[T1]{fontenc}
\usepackage{lmodern}
\usepackage[margin=1in]{geometry}
\usepackage{microtype}

\usepackage{amsmath, amssymb, amsthm}
\usepackage{mathtools}
\usepackage{thm-restate}

\usepackage{graphicx}
\graphicspath{{figures/}}

\usepackage[justification=centering]{caption}
\usepackage[dvipsnames]{xcolor}

\usepackage{tikz}

\usepackage{algorithm}
\usepackage[noend]{algorithmic}

\usepackage[
  backend=biber,
  style=numeric,
  citestyle=numeric-comp,
  sorting=nyt,
  sortcites=true,
  maxbibnames=99,
  giveninits=false,
  doi=true,
  isbn=true,
  url=true
]{biblatex}

\theoremstyle{plain}
\newtheorem{theorem}{Theorem}
\newtheorem{lemma}[theorem]{Lemma}

\theoremstyle{definition}

\theoremstyle{remark}
\newtheorem{remark}[theorem]{Remark}

\usepackage[
  colorlinks=true,
  linkcolor=red,
  citecolor=blue!90!black,
  urlcolor=black
]{hyperref}

\usepackage[capitalise,noabbrev,nameinlink]{cleveref}

\crefname{algorithm}{Algorithm}{Algorithms}
\crefname{property}{Property}{Properties}

\crefname{appendix}{Appendix}{Appendices}
\Crefname{appendix}{Appendix}{Appendices}

\newenvironment{prosealgorithm}[1]{%
  \begin{list}{}{%
    \setlength{\leftmargin}{2em}%
    \setlength{\rightmargin}{2em}%
    \setlength{\topsep}{0.6\baselineskip}%
    \setlength{\partopsep}{0pt}%
    \setlength{\parsep}{0.2\baselineskip}%
    \setlength{\itemsep}{0pt}%
  }%
  \item[]\textbf{\textit{#1}:}\ \ignorespaces
}{%
  \end{list}%
}

\newcommand{\BinPacking}{\textsc{Bin Packing}}
\newcommand{\DynamicBinPacking}{\textsc{Dynamic Bin Packing}}
\newcommand{\TimelyBinPacking}{\textsc{Timely Bin Packing}}
\newcommand{\FractionalTimelyBinPacking}{\textsc{Fractional Timely Bin Packing}}

\DeclareMathOperator{\OPT}{OPT}
\DeclareMathOperator{\ALG}{Thrifty\text{-}Greedy}
\DeclareMathOperator{\Greedy}{Greedy}
\DeclareMathOperator{\DiscreteGreedy}{Discrete\text{-}Greedy}

\DeclareMathOperator{\PackAtDeadline}{Pack\text{-}at\text{-}Deadline}

\title{Online Timely Bin Packing Revisited:\\
A Simpler Algorithm with Optimality in Special Cases}

\author{
  Pei-Lin Chiang\thanks{
    Department of Computer Science, National Tsing Hua University, Taiwan. 
    \href{mailto:ivylingchiang@gmail.com}{\texttt{ivylingchiang@gmail.com}}
  }
  \and 
  Wing-Kai Hon\thanks{
    Department of Computer Science, National Tsing Hua University, Taiwan. 
    \href{mailto:wkhon@cs.nthu.edu.tw}{\texttt{wkhon@cs.nthu.edu.tw}}
  }
  \and 
  Ying-Chia Huang\thanks{
    Department of Computer Science, National Tsing Hua University, Taiwan. 
    \href{mailto:ereal3478@gmail.com}{\texttt{ereal3478@gmail.com}}
  }
  \and 
  Li-Tzu Yang\thanks{
    Department of Computer Science, National Tsing Hua University, Taiwan. 
    \href{mailto:breton0429@gmail.com}{\texttt{breton0429@gmail.com}}
  }
}

\date{}

\hypersetup{
  pdftitle={
    Online Timely Bin Packing Revisited: A Simpler Algorithm with Optimality in Special Cases
  },
  pdfauthor={
    Pei-Lin Chiang,
    Wing-Kai Hon,
    Ying-Chia Huang,
    and Li-Tzu Yang
  }
}

\begin{document}

\maketitle

\begin{abstract}

Online \textsc{Timely Bin Packing} is a variant of classical online bin packing in which items arrive over time with sizes and deadlines. Packing an item may be deferred until its deadline, but once a collection of items is packed into a bin, the bin is closed and cannot be used again. The objective is to minimize the total number of bins used. The problem was introduced by Khattar in 2023. In 2025, Aminian proposed Discrete-Greedy, a deterministic online algorithm with a competitive ratio of exactly $2$, which remains the best-known upper bound. Its analysis is particularly simple; however, the algorithm requires maintaining a fractional simulation internally and may leave usable space in some bins unfilled.

We propose Thrifty-Greedy, a simple deterministic online algorithm that does not require a fractional simulation. We prove that its competitive ratio is exactly $2$, matching the best-known upper bound. At the same time, the algorithm automatically achieves the optimal competitive ratio on two natural classes of inputs: it is $1$-competitive when all items have the same size, and $1.5$-competitive when every item has size in $\{1/3,2/3\}$. In contrast, Discrete-Greedy has a competitive ratio of at least $2$ on both classes.

\end{abstract}

\section{Introduction}
\label{sec:introduction}

In classical online \BinPacking{}, items arrive one by one, with each item $p_i$ having a size $s_i \in (0,1]$.
Upon arrival, each item must be assigned immediately and irrevocably to a unit-capacity bin, and the objective is to minimize the total number of bins used.
Note that the model has no explicit notion of time beyond the order of arrivals.

A classical way to introduce explicit time constraints is \DynamicBinPacking{}, introduced by Coffman~et~al.\ in 1983~\cite{coffman1983dynamic}.
In this model, each item has an arrival time and a departure time.
An arriving item must be assigned immediately to a bin; when the item later departs, the space it occupied becomes available again, while the bin itself remains in the system.
This models persistent resources whose capacity can be reused over time, such as cloud servers that remain available while jobs arrive and depart.

In this paper, we study another way to introduce time constraints: online \TimelyBinPacking{}.
This problem is introduced by Khattar in 2023~\cite{khattar2023online}.
Here, an arriving item has a deadline, and the algorithm may defer packing it until any time no later than that deadline.
When several items are packed together, the entire bin is closed at that time and cannot be used again.
Thus, unlike in \DynamicBinPacking{}, items do not depart individually and leave reusable space; instead, the bin itself leaves the system.

A motivating example comes from logistics.
Items arrive at a warehouse over time and may be stored temporarily before dispatch.
Each item has a deadline by which it must be loaded onto a truck or container, and each truck has a fixed capacity.
Several items can be dispatched together only if they all fit in the truck and are present at the warehouse when the truck departs.
Because trucks are costly, the objective is to minimize the number of dispatched trucks.
For simplicity, the model assumes that pending items can be stored at the warehouse without limit and at no cost; only the number of dispatched trucks is counted.

Let us now formally define online \TimelyBinPacking{}.

\subsection{Problem Definition}
\label{sec:problem_def}

We are provided with an unlimited number of bins, each of capacity $1$.
The input is a finite sequence $I=(p_1,p_2,\ldots,p_n)$ of items.
Each item $p_i$ is associated with three parameters: an arrival time $a_i$, a deadline $d_i$, and a size $s_i \in (0,1]$.
The items arrive online, meaning that the parameters of an item $p_i$ are revealed only when it arrives at time $a_i$.
We assume that $a_i \le d_i$ for every item $p_i$, and call $[a_i,d_i]$ the \emph{interval} of $p_i$.
We index the items in nondecreasing order of arrival time, so that $a_1 \le a_2 \le \cdots \le a_n$.
Whenever items have the same arrival time or deadline, ties are broken by item index.

For each item $p_i$, we need to pack it into some bin at some packing time $t_j$ such that $t_j \in [a_i,d_i]$.
Unlike in classical online \BinPacking{}, where an item must be packed immediately upon arrival, in online \TimelyBinPacking{}, the decision of when to pack an item may be deferred from its arrival until any time no later than its deadline.
Yet, once the decision is made, it cannot be changed.
An item that has arrived but has not yet been packed is called \emph{pending}.

Formally, given an algorithm $\mathcal{A}$ and an online input sequence $I$, the output of $\mathcal{A}$ is a collection of bins $O(\mathcal{A},I)=\{b_1,b_2,\ldots,b_m\}$ satisfying the following conditions:
\begin{enumerate}
    \item Each bin $b_j \in O(\mathcal{A},I)$ is associated with a packing time $t_j$.
    \item Each item $p_i$ is assigned to exactly one bin $b_j$ whose packing time satisfies $t_j \in [a_i,d_i]$.
    \item For every bin $b_j$, the total size of the items assigned to $b_j$ is at most $1$.
\end{enumerate}

We measure the performance of an online algorithm using standard \emph{competitive analysis}.
The cost of an algorithm $\mathcal{A}$ on input $I$, denoted by $\mathcal{A}(I)$, is the number of bins in its output, i.e., $\mathcal{A}(I)=|O(\mathcal{A},I)|$.
Let $\OPT$ denote an optimal offline algorithm that knows the entire input in advance and, for every input $I$, uses the minimum number of bins.
For $c \ge 1$, we say that an online algorithm $\mathcal{A}$ is \emph{$c$-competitive} if $\mathcal{A}(I) \le c\,\OPT(I)$ for every input $I$.
The \emph{competitive ratio} of $\mathcal{A}$ is the infimum of all such values $c$.
Our goal is to design an online algorithm $\mathcal{A}$ that minimizes this ratio.

\subsection{Prior Work}
\label{sec:prior_work}

Khattar introduced online \TimelyBinPacking{} and proved that no deterministic online algorithm can have a competitive ratio below $1.2$~\cite[Theorem~3]{khattar2023online}.
Khattar also proposed the \emph{Pack-at-Deadline} algorithm and claimed that it is $2$-competitive~\cite[Theorem~1]{khattar2023online}.
Aminian subsequently showed that this claim is incorrect by proving that $\PackAtDeadline$ has a competitive ratio of at least $2.5$~\cite[Theorem~4.2.5]{aminian2025algorithms}.

Aminian also improved both the general lower and upper bounds.
Aminian proved a lower bound of $1.5$ for the competitive ratio of any deterministic online algorithm, even on inputs containing only items of sizes in $\{1/3,2/3\}$~\cite[Theorem~4.5.12]{aminian2025algorithms}.
On the algorithmic side, Aminian introduced \emph{Discrete-Greedy}, a $2$-competitive deterministic online algorithm~\cite[Theorem~4.2.10]{aminian2025algorithms}.
A matching lower bound shows that $\DiscreteGreedy$ is no better than $2$-competitive, so this analysis is tight~\cite[Theorem~4.2.11]{aminian2025algorithms}.
Aminian's thesis also studies several other variants, including uniform-size items, unit-length intervals, integer arrival times, randomized algorithms, advice, and the offline problem.

Because $\DiscreteGreedy$ is the main benchmark for our work, we next briefly review its definition and analysis.

\paragraph{The fractional relaxation.}
The definition and analysis of $\DiscreteGreedy$ are based on a fractional relaxation of the problem.
In online \FractionalTimelyBinPacking{}, an item may be divided into arbitrarily many portions (intuitively, thinking of items as liquids), and different portions may be packed in different bins, as long as every portion is packed at a time within the original item's interval.
Let $\OPT_f(I)$ denote the optimal offline cost in this fractional model.
Since every feasible integral packing is also a feasible fractional packing, we have
\[
\OPT_f(I) \le \OPT(I).
\]

Aminian first designed a simple earliest-deadline-first algorithm that produces an optimal packing in the fractional setting.

\begin{prosealgorithm}{Greedy}
At any time $t$ when at least one pending item reaches its deadline, open a bin and consider all currently available pending items in nondecreasing order of deadline, breaking ties by item index.
Pack the items in this order until no item remains or the next item would exceed the remaining capacity.
In the latter case, split that item, place exactly the portion needed to fill the bin, and leave the remaining portion pending.
Repeat at time $t$ while some pending item has deadline $t$.
\end{prosealgorithm}

A standard exchange argument shows that $\Greedy$ is optimal even among offline fractional packings~\cite[Lemma~4.2.9]{aminian2025algorithms}.
The argument considers the first bin on which an optimal fractional packing differs from $\Greedy$ and exchanges portions of later-deadline items for equal-sized portions of earlier-deadline items, without increasing the number of bins.
Repeating the exchange transforms the optimal packing into the one produced by $\Greedy$.
Consequently, for every input $I$,
\begin{equation}
    \Greedy(I)=\OPT_f(I)\le \OPT(I).
    \label{eq:greedy-fractional-optimal}
\end{equation}

Aminian then used $\Greedy$ to design a simple $2$-competitive online algorithm for the integral problem.

\begin{prosealgorithm}{Discrete-Greedy}
Simulate $\Greedy$.
For each fractional bin produced by $\Greedy$, pack the items that occur wholly in that bin into one integral bin, omitting fractional portions of items that were split elsewhere.
If $\Greedy$ ends the fractional bin by splitting an item $p$, open an additional dedicated bin containing the whole item $p$ and close it at the same time.
Future fractional portions of $p$ are omitted from the simulated packing.
\end{prosealgorithm}

Thus, each bin of $\Greedy$ projects to at most two bins of $\DiscreteGreedy$.
Together with~\eqref{eq:greedy-fractional-optimal}, this immediately gives
\[
    \DiscreteGreedy(I) \le 2\,\Greedy(I)
    =2\,\OPT_f(I)
    \le 2\,\OPT(I).
\]
The resulting analysis is very simple and straightforward.
However, the dedicated-bin rule is deliberately conservative: an item split by the fractional simulation is placed alone, even when it could share a bin with other currently available items.
Moreover, the integral algorithm must maintain the state of the fractional simulation, including any remaining fractional portions.

\subsection{Our Contributions}
\label{sec:contributions}

The definition of $\DiscreteGreedy$ naturally raises the following question: if the next item in earliest-deadline order does not fit in the current bin, why should the algorithm immediately give this item its own bin? 
A more natural choice is to simply leave that item pending and continue trying to fill the remaining space of the current bin with other items.
In this paper, we propose an algorithm based on this simple idea, which we call \emph{Thrifty-Greedy}.
Informally, whenever a pending item reaches its deadline, $\ALG$ opens a bin for an earliest-deadline pending item and then scans the other pending items in earliest-deadline order, packing every item that fits and skipping those that do not.
This simple heuristic avoids opening a dedicated bin for the skipped item and also eliminates the need to simulate a fractional packing internally.
The formal definition is given in~\cref{sec:algorithm}.

We analyze $\ALG$ in the general setting and in two natural restricted settings.
Our results show that this simple heuristic matches the competitive ratio of $\DiscreteGreedy$ in the general setting, while attaining the optimal competitive ratio on two restricted classes for which $\DiscreteGreedy$ is suboptimal.
Notably, these two restricted guarantees are ``opportunistic'': $\ALG$ does not need to know in advance that the input belongs to either restricted class.
The same general-purpose algorithm automatically attains the optimal competitive ratio whenever the input belongs to one of them.

\paragraph{General inputs.}
In~\cref{sec:general_case}, we show that $\ALG$ is $2$-competitive, matching $\DiscreteGreedy$.
However, $\ALG$ is not better than $\DiscreteGreedy$ in the general setting: we also give a lower-bound construction showing that $\ALG$ is no better than $2$-competitive.

The projection argument used for $\DiscreteGreedy$ no longer applies directly, because bins of $\ALG$ do not correspond locally to bins of $\Greedy$.
Nevertheless, we show that we can still prove the analogous global inequality $\ALG(I) \le 2\,\Greedy(I)$ by strong induction on the number of items in $I$.
At a high level, we show that we can always select two bins produced by $\ALG$ and remove their entire contents from the input so that the removed items contain the entire contents of at least one bin produced by $\Greedy$.
Thus, the number of bins used by $\ALG$ decreases by two, while the number of bins used by $\Greedy$ decreases by at least one.
Applying the induction hypothesis to the remaining input yields the desired factor of $2$.

\paragraph{Uniform-size items.}
In~\cref{sec:uniform_case}, we show that $\ALG$ is $1$-competitive when all items have the same size, i.e., it uses the same number of bins as any optimal offline algorithm.
In fact, on uniform-size inputs, $\ALG$ follows the same strategy as Aminian's \emph{Greedy-on-Deadline} algorithm, which was defined specifically for this class of inputs and was shown to be optimal~\cite[Observation~4.5.1]{aminian2025algorithms}.
For completeness, we provide a proof using a standard exchange argument.
In contrast, $\DiscreteGreedy$ has a competitive ratio of at least $2$ even on uniform-size inputs; see~\cref{appendix:Discrete_Greedy_LB}.

\paragraph{Items of sizes in $\{1/3,2/3\}$.}
In~\cref{sec:special_case}, we show that $\ALG$ is $1.5$-competitive when all items have sizes in $\{1/3,2/3\}$.
This ratio is optimal for the class, since Aminian proved a lower bound of $1.5$ for any deterministic online algorithm, even on inputs containing only items of these two sizes~\cite[Theorem~4.5.12]{aminian2025algorithms}.
Again, in contrast, $\DiscreteGreedy$ has a competitive ratio of at least $2$ on this restricted class of inputs; see~\cref{appendix:Discrete_Greedy_LB}.

The proof is rather involved, although the high-level idea is simple.
Intuitively, we want to show that the bins produced by $\ALG$ are $2/3$-full on average.
A $2/3$-full or full bin already has sufficient utilization, and pairing a $1/3$-full bin with a full bin also gives an average load of $2/3$.
Thus, the problematic bins are the remaining unpaired $1/3$-full bins.
If there are $x$ such bins, we would like to argue that an optimal packing must also contain at least $x$ $1/3$-full bins.
Unfortunately, establishing such a lower bound directly appears difficult.

We therefore give an input-conversion procedure that successively modifies the intervals of some items, converting $I$ into an input $I'$ for which the desired lower bound is easier to establish.
The complication is that these modifications may also change the packing produced by $\ALG$.
We must therefore ensure throughout the conversion that $\ALG$ does not use fewer bins on $I'$ than on $I$.
Maintaining this guarantee while creating the structure needed for the lower bound is the main reason the proof becomes technical.

\subsection{Other Related Work}
\label{sec:other_related_work}

Interestingly, the offline version of \TimelyBinPacking{} has been studied from a graph-theoretic perspective.
Represent each item $p_i$ by the interval $[a_i,d_i]$ with weight $s_i$.
A set of items can share a bin exactly when their intervals have a common intersection and their total weight is at most $1$.
Equivalently, a feasible bin corresponds to a clique of total weight at most $1$ in the associated weighted interval graph.
The offline problem is therefore equivalent to partitioning a weighted interval graph into the minimum number of bounded-weight cliques.

Chen et al.~\cite{chen2007clique} studied this problem under the name \emph{minimum clique partition with constrained bounds in weighted interval graphs}.
They proved that it is strongly NP-hard~\cite[Theorem~1]{chen2007clique} and cannot be approximated within a factor of $3/2-\varepsilon$, for any $\varepsilon>0$, unless $\mathsf{P}=\mathsf{NP}$~\cite[Theorem~2]{chen2007clique}.
Myung~\cite{myung2008clique} later gave a $2$-approximation algorithm for the case of arbitrary weights, and an exact algorithm for the case of identical weights.

Thus, assuming $\mathsf{P}\neq\mathsf{NP}$, the currently best-known lower and upper bounds on the optimal polynomial-time approximation ratio for \TimelyBinPacking{} are $1.5$ and $2$, respectively.
We remark that the same numerical gap remains in the online setting: the currently best-known lower and upper bounds on the optimal competitive ratio are also $1.5$ and $2$, respectively.

\section{The Proposed Algorithm}
\label{sec:algorithm}

In this section, we formally define $\ALG$. Recall that at any time $t$, an item is \emph{pending} if it has arrived but has not yet been packed.

\begin{prosealgorithm}{Thrifty-Greedy}
At any time $t$ at which at least one pending item reaches its deadline, repeat the following procedure until no pending item has deadline $t$.
\begin{enumerate}
    \item Choose a pending item $p$ with the earliest deadline (i.e., deadline $t$), breaking ties by item index. Open a new bin $b$, assign it packing time $t$, and pack $p$ into $b$. We call $p$ the \emph{trigger item} of $b$.
    \item Scan the remaining pending items in nondecreasing order of deadline, again breaking ties by item index. Whenever the current item fits in the remaining capacity of $b$, pack it into $b$; otherwise, skip it and leave it pending. After every pending item has been considered, close and output $b$.
\end{enumerate}
\end{prosealgorithm}

For an input sequence of items $I$, let $O$ denote the corresponding output sequence of bins.
For each bin $b \in O$, let $w(b)$ denote the total size of all items packed in $b$.
One reason the proposed algorithm works well is the following lemma.

\begin{lemma} \label{lem:trigger}
Let $b$ and $b'$ be two distinct bins in $O$, and let $p$ and $p'$ denote their trigger items, 
respectively. If $w(b) + w(b') \leq 1$, then the intervals of $p$ and $p'$
do not overlap.  
\end{lemma}
\begin{proof}
Suppose, for contradiction, that the intervals of $p$ and $p'$ overlap. 
Without loss of generality, assume the deadline of $p$ is within the interval of $p'$.
Then, by the time $b$ is opened at the deadline of $p$, the item $p'$ is pending.
Also, $w(b) + (\text{size of $p'$}) \leq w(b) + w(b') \leq 1$, so by our algorithm, $p'$ would be packed
into~$b$.  Yet, $p'$ is packed in another bin $b'$, a contradiction.
\end{proof}

\section{Performance Analysis}
We analyze $\ALG$ in three settings: uniform-size items, general inputs, and items with sizes in $\{1/3,2/3\}$.
The subsections are presented in increasing order of the technical complexity of their analyses.

\subsection{Uniform-Size Items}
\label{sec:uniform_case}

Consider the special case where every item in the input $I$ has the same size.
Let $k \in \mathbb{N}$ be the maximum number of items that a bin can hold.  
We show that $\ALG$ uses the same number of bins as any optimal offline algorithm in this case.

Recall that our proposed algorithm is a \emph{greedy} algorithm, where we will always (i) pick
an item $p$ with the earliest deadline, say $d_p$, to be packed in some bin $b$, 
(ii) set $b$'s packing time to $d_p$, and (iii) add other items to $b$ (prioritized by their
deadlines, the earlier the better) as many as possible.
To show that our algorithm is optimal, we use the standard exchange argument for greedy algorithms.
We first show that
such a \emph{greedy choice} is consistent with the choice of some optimal algorithm.
Then, inductively, all the greedy choices we have made will be the same as the choices 
of some optimal algorithm.

Let $\OPT$ be an optimal algorithm.  
Let $p$ denote an item with the earliest deadline in input $I$, and $d_p$ be its deadline.
Let $Q$ be the set of items that are available at time $d_p$ except item $p$.
That is, the arrival time of each item in $Q$ is on or before $d_p$.
Also, let $Q' \subseteq Q$ be the set of $\min(k-1, |Q|)$ items in $Q$ whose deadlines are the earliest,
such that these items are exactly those packed in the same bin with $p$ in our greedy algorithm.

Consider how $\OPT$ would pack $p$.   Since $p$ has the earliest deadline, without loss of generality, 
we can assume that $\OPT$ sets the packing time of the bin containing $p$, say $b^*$, as~$d_p$.  
We have the following two scenarios:

\begin{itemize}
\item If $\OPT$ packs $p$ with all the items of $Q'$, done.

\item Else, there exists some item $q' \in Q'$ not packed with $p$ by $\OPT$.
      If the bin $b^*$ containing $p$ still has vacant space for $q'$, 
      we modify the choice of $\OPT$ by moving $q'$ into $b^*$.  
      This change does not increase the number of bins used, so the resulting packing is also optimal; denote the corresponding optimal algorithm by $\OPT'$.    
      Repeating the above process, we can put more items of $Q'$ into $b^*$,
      while maintaining optimality.
      
      In the end, either all items of $Q'$ are now in $b^*$, in which case we are done,
      or, there exists some $q' \in Q'$ in some bin $b$ other than $b^*$.
      For the latter case, there must be some $q \not\in Q'$ packed in $b^*$,
      and we modify the choices of the optimal algorithm again, 
      by exchanging $q$ and $q'$ in their corresponding bins.
      Note that such a change is valid, for $q'$ has a deadline earlier than or equal to $q$,
      so that $q$ can be moved to $b$ (as its packing time is on or before~$q'$),
      while $q'$ can be moved to $b^*$ (as its deadline is on or after $d_p$).
      Because this exchange does not increase the number of bins used, the resulting packing is still optimal.
      Repeating the above exchange process, we can put all items of $Q'$ into $b^*$,
      while maintaining optimality.
\end{itemize}

The above discussion shows that there exists some optimal algorithm that makes
exactly the same choice as we do, for packing items along with $p$.
Inductively, there exists some optimal algorithm that makes exactly the same choices
as we do, for packing each of the items.
This gives the following theorem.

\begin{restatable}{theorem}{ThmUniformOptimality}
\label{thm:uniform_optimality}
When all items have the same size, $\ALG(I)=\OPT(I)$ for every such input $I$.
\end{restatable}

\subsection{General Inputs}
\label{sec:general_case}

We now consider unrestricted item sizes and prove that $\ALG$ is $2$-competitive, with a matching lower bound of $2$.

We first prove the upper bound. Recall from
\eqref{eq:greedy-fractional-optimal} that the fractional algorithm $\Greedy$
reviewed in~\cref{sec:prior_work} satisfies
\[
    \Greedy(I)=\OPT_f(I)\le \OPT(I)
\]
for every input $I$. 
Therefore, it suffices to establish the following comparison between $\ALG$ and $\Greedy$.

\begin{lemma}
\label{lem:general-comparison}
For every input $I$, $\ALG(I) \le 2\, \Greedy(I)$.
\end{lemma}
\begin{proof}
We prove the lemma by strong induction on the length of the input $I$.

\textbf{Base Case:} When $|I| = 1$, the input consists of a single item; both $\ALG$ and $\Greedy$ will pack this item into one bin. Thus, $\ALG(I) = \Greedy(I) = 1$, and the lemma holds.

\textbf{Inductive Step:} Assume the lemma holds for all inputs of length at most $k$, for some $k \in \mathbb{N}$. Consider an input $I$ with $|I| = k+1$. Let us compare the sequences of bins produced by $\ALG$ and $\Greedy$ on input $I$, indexed by their packing orders. We scan the two sequences from left to right to find the first index $i$, where the contents of their $i$th bins differ. If no such index exists, the packings are identical, implying $\ALG(I) = \Greedy(I)$, so the lemma holds. Otherwise, suppose the packings differ at the $i$th bin. Since both algorithms pack pending items by increasing order of deadlines, they consider the same set of items in the same order. Therefore, the difference must arise because $\Greedy$ can split an item that fits only partially, whereas $\ALG$ cannot.

Specifically, both algorithms pack a set of items $C^*$ into the $i$th bin. The next item in the priority order, let us call it $A$, does not fit entirely into the remaining space.  Then, the corresponding actions of the two algorithms are as follows.
\begin{itemize}
    \item $\ALG$: Cannot split $A$, so it skips $A$ and continues scanning for other items that fit. Let $D^*$ be the set of subsequent items that $\ALG$ packs into the $i$th bin. Thus, the $i$th bin of $\ALG$ contains $C^* \cup D^*$.
    \item $\Greedy$: Splits $A$ into two pieces, $a_1$ and $a_2$. It packs $a_1$ into the $i$th bin, making it full. The remainder $a_2$ is left for subsequent bins. Thus, the $i$th bin of $\Greedy$ contains $C^* \cup \{a_1\}$.
\end{itemize}

Since $\ALG$ skips $A$ at the $i$th bin, it must pack $A$ into some later bin, say, the $j$th bin (where $j > i$). Let $B^*$ denote the set of other items packed alongside $A$ in this $j$th $\ALG$ bin. Let us now construct a smaller input $I'$ by removing all items contained in the $i$th and $j$th $\ALG$ bins from the original input $I$. Specifically, let $I' = I \setminus (C^* \cup D^* \cup \{A\} \cup B^*)$.

Since $C^* \cup D^*$ and $\{A\} \cup B^*$ constitute the complete contents of two bins in $\ALG$, removing these items from the input does not affect the packing of the remaining bins (they are packed identically as before). Therefore, the total number of bins used by $\ALG$ on $I'$ decreases by exactly $2$. Notice that the length of the new input is strictly shorter than $I$ (i.e., $|I'| \le k$), hence by the induction hypothesis we have
\[
\ALG(I)\ =\ 2 + \ALG(I')\ \le\ 2 + 2 \cdot \Greedy(I').
\]

Next, consider the behavior of $\Greedy$. Suppose we modify the original input $I$ by replacing item $A$ with its two split pieces, $a_1$ and $a_2$ (where both inherit the same arrival time and deadline as $A$). Since $\Greedy$ operates in the fractional setting and the packing order depends only on arrival times and deadlines, this replacement does not change the original packing. Also, since $C^* \cup \{a_1\}$ constitutes the complete content of one bin in the $\Greedy$ packing, we can apply a removal argument similar to the one used for $\ALG$. Thus, removing those items from the input does not affect the packing of the remaining bins, so that:
\begin{align*}
\Greedy(I)\ &=\ \Greedy((I \setminus \{A\}) \cup \{a_1, a_2\})\\
                   &=\ 1 + \Greedy(I' \cup D^* \cup \{a_2\} \cup B^*).
\end{align*}

Our final step is to show that
$\Greedy(I' \cup D^* \cup \{a_2\} \cup B^*) \geq \Greedy(I')$.
To establish this, we use the following notion of ``monotonicity''.
We say that an input $Y$ \emph{dominates} an input $X$ if $Y$ can be obtained from $X$ by adding items, increasing their sizes, and/or restricting their intervals.
An algorithm $\mathcal{A}$ is \emph{monotone} if
$\mathcal{A}(X) \leq \mathcal{A}(Y)$ whenever $Y$ dominates $X$.
Intuitively, adding items, making them larger, or giving them narrower intervals should not make the packing easier.

Although one might expect this property to hold generally, many classical bin packing algorithms are not monotone~\cite{murgolo1984on,murgolo1988anomalous}.
Moreover, the standard counterexample for the classical First-Fit heuristic from~\cite{murgolo1988anomalous} can be adapted to show that $\ALG$ is not monotone for online \TimelyBinPacking{}.

Consider an input $X$ dominated by $Y$. 
Notice that any fractional solution for $Y$ is also a feasible fractional solution for $X$, 
so we have $\OPT_f(X) \le \OPT_f(Y)$.
Since $\Greedy$ is optimal, we have $\Greedy(X) = \OPT_f(X) \le \OPT_f(Y) = \Greedy(Y)$.
Thus, $\Greedy$ is monotone. 

Next, we see that $I'$ is dominated by $I' \cup D^* \cup \{a_2\} \cup B^*$. 
By the monotonicity of $\Greedy$, we have $\Greedy(I' \cup D^* \cup \{a_2\} \cup B^* ) \ge \Greedy(I')$. 
Combining the results in the above discussion, we have
\begin{align*}
\ALG(I)\ &\le\ 2 + 2 \cdot \Greedy(I') \\
                &\le\ 2 + 2 \cdot \Greedy(I' \cup D^* \cup \{a_2\} \cup B^* )\\ 
                &=\ 2 \cdot \Greedy(I).
\end{align*}
Thus, the lemma holds for $|I| = k+1$. This completes the inductive step.
\end{proof}

We have thus proven that $\ALG$ is $2$-competitive.
We now show that the above competitive analysis is indeed tight, i.e., $\ALG$ is not better than $2$-competitive.

Consider the following input.
Let $N$ be a large integer and let $0<\varepsilon\le 1/N$. 
At time $t=0$, $N$ items of size $1 - \varepsilon$ arrive with an arbitrarily late deadline. At each time $t=1,2,\dots,N$, two items of size $\varepsilon$ arrive: one with an immediate deadline at $t$, and the other with a deadline at $N$. $\ALG$ packs the two $\varepsilon$-items together in one bin at each $t=1,2,\dots,N$, and uses another $N$ separate bins for the large ($1 - \varepsilon$)-items, using $2N$ bins in total. In contrast, $\OPT$ can pack the immediate $\varepsilon$-item with a ($1 - \varepsilon$)-item at each $t = 1, 2, \dots, N$, and pack all $N$ remaining $\varepsilon$-items into a single bin at time $t=N$, using only $N+1$ bins. As $N \to \infty$, this shows a lower bound of~$2$ for the competitive ratio of $\ALG$.

Summarizing, we have the following theorem.

\begin{restatable}{theorem}{ThmGeneralCR}
\label{thm:general_CR}
In the general setting, $\ALG$ is $2$-competitive, which is tight.
\end{restatable}

\subsection{Items of Sizes in \texorpdfstring{$\{1/3,2/3\}$}{\{1/3, 2/3\}}}
\label{sec:special_case}

We now consider the special case where every item has size in $\{1/3,2/3\}$.
We show that $\ALG$ is $1.5$-competitive in this setting, which is optimal.

Since every item has size $1/3$ or $2/3$, we classify the bins used by an algorithm into 
\emph{small}, \emph{medium}, and \emph{large} bins, which respectively are $1/3$-full, $2/3$-full, and full.  
More precisely, let
\begin{itemize} 
  \item $x^*, y^*, z^*$ be the number of small, medium, and large bins used by an optimal offline algorithm $\OPT$ on the original input $I$;
  \item $x, y, z$ be the number of small, medium, and large bins used by our algorithm $\ALG$ on the original input $I$;
  \item $x', y', z'$ be the number of small, medium, and large bins used by our algorithm $\ALG$ on the modified input $I'$.
\end{itemize}

At a high level, we would like to show that the bins produced by $\ALG$ are $2/3$-full on average, which would give a competitive ratio of $1.5$.
A medium or large bin already has load at least $2/3$, while pairing a small bin with a large bin also gives an average load of $2/3$.
Thus, the problematic bins are the remaining unpaired small bins.
We would like to show that if $\ALG$ performs badly on such a bin, then we can associate it with a distinct small bin where $\OPT$ also performs badly.
This would give the desired bound on the competitive ratio.
However, the packings produced by $\ALG$ and $\OPT$ may be very different, making such a direct association difficult.

In~\cref{sec:uniform_case}, we handled a similar comparison using a standard exchange argument: we successively modified the packing produced by $\OPT$ so that it remained as \emph{good}, while making its choices agree with those of $\ALG$.
Here, we take an orthogonal approach.
Instead of modifying either algorithm's packing, we keep the algorithms fixed and successively modify the input $I$ into an input $I'$ so that $\ALG$ remains as \emph{bad} on the modified input, while making it easier to associate the remaining unpaired small bins of $\ALG$ on $I'$ with small bins of $\OPT$ on the original input $I$.
By repeatedly applying the conversion, we can compare $\ALG$ on the final modified input $I'$ with $\OPT$ on the original input $I$, and consequently obtain the desired competitive analysis.

In~\cref{sec:input_conversion}, we present the input-conversion algorithm and prove its correctness.
Specifically, the algorithm converts any input $I$ into another input $I'$ such that the following three properties hold.
\begin{itemize}
   \item We use the same set of items in $I$ and in $I'$, except some items have more ``restrictive'' arrival times and deadlines in $I'$. Since the total size of items remains unchanged, $x^* + 2y^* + 3z^* = x + 2y + 3z = x' + 2y' + 3z'$.
   \item After pairing small bins with large bins used by $\ALG$ on $I'$, the number of remaining unpaired small bins is at most the number of small bins used by $\OPT$ on $I$.
      In particular, $x' - z' \leq x^*$.
   \item Our algorithm $\ALG$ performs as badly on $I'$ as on $I$. That is, $\ALG(I) \leq \ALG(I')$. 
\end{itemize}

\begin{remark}
The first property does not imply the third.
Recall from~\cref{sec:general_case} that $I'$ dominates $I$, but $\ALG$ is not monotone.
Therefore, it is not guaranteed that $\ALG(I)\leq \ALG(I')$.
\end{remark}

For now, suppose that we have such an input-conversion algorithm.
Then, we have the following theorem.

\begin{theorem}
When every item has size in $\{1/3, 2/3\}$, $\ALG$ is $1.5$-competitive, which is optimal.
\end{theorem}
\begin{proof}
We adopt the notation defined above. 
By combining the equality from the first property with the inequality from the second property, we obtain 
\[
2x' + 2y' + 2z'\ \leq\ 2x^* + 2y^* + 3z^*.
\]
Consequently, we have 
\[
\ALG(I')\ =\ x' + y' + z'\ \leq\ 1.5\,(x^* + y^* + z^*)\ =\ 1.5\cdot \OPT(I).
\]
Combining this with the third property, we obtain 
\[
\ALG(I)\ \leq\ \ALG(I')\ \leq\ 1.5\cdot \OPT(I)
\]
as desired. 
Furthermore, our algorithm is optimal, since Aminian proved a lower bound of $1.5$ for any deterministic online algorithm, even on inputs containing only items of sizes in $\{1/3, 2/3\}$.
This completes the proof.
\end{proof}

\section{Input Conversion for the \texorpdfstring{$\{1/3,2/3\}$}{\{1/3, 2/3\}} Case}
\label{sec:input_conversion}

In this section, we will give details on the input conversion mentioned in \cref{sec:special_case}
to change the input $I$ into another sequence $I'$ satisfying certain properties.  
Note that the sequence $I'$ is only for analysis purposes, so in the construction of such a sequence,
we will assume that some optimal offline algorithm $\OPT$ is known, and so is its corresponding
output $O^*$ on the input $I$.  

We consider the special case where each item in the input has size either $1/3$ or $2/3$.
Here, we adopt the notation defined in \cref{sec:special_case}. 
The following are the three properties that $I'$ will be shown to satisfy in this section:

\begin{itemize}
   \item $I$ and $I'$ use the same set of items,
         except some items may have more ``restrictive'' arrival times and deadlines in $I'$.
   \item $x' - z' \leq x^*$.  
   \item $\ALG(I) \leq \ALG(I')$.         
\end{itemize}

\subsection{The Conversion Algorithm}
The formal definition is given in \cref{alg:special_case}.
In the algorithm, we first fix an optimal offline algorithm $\OPT$ and obtain its output $O^*$ on $I$ for reference (Line 1 and Line 2). 
Then, we initialize $I'$ as $I$ (Line 3), and iteratively update $I'$ by modifying the arrival time and deadline of some selected item.
Every time $I'$ is updated, we will rerun our $\ALG$ on $I'$ to obtain the corresponding output $O'$ (Line 4 and Line 10).   

The algorithm maintains a time $t$ which is initialized as the latest deadline of all items in $I$ (or the initial $I'$).
This time $t$ will be updated iteratively (Line 9), each time moving to an earlier moment.
Based on $t$, we select the latest small bin in $O'$ with packing time before $t$, and also its trigger item $C$ (Line 7 and Line 8), 
and then modify $t$ and some other item's arrival time or deadline based on $C$, $O'$ and $O^*$(Line 9).  
Consequently, the sequence $I'$ is updated.

When no small bins with packing time before $t$ are found, the conversion algorithm terminates,
in which case the sequence $I'$ is obtained (Line 11).

Based on $C$, $O'$ and $O^*$, the modification in Line 9 is made according to the following three cases:
\begin{itemize}
\item Case 1:  In $O^*$, the item $C$ is also in a small bin.
\item Case 2:  In $O'$, there exists some other bin packed in the interval of $C$.
\item Case 3:  Neither Case 1 nor Case 2 applies.
\end{itemize}
The intuition of this case division is that in the first two cases, we do not even need to modify $I'$, since the output $O'$ is already good enough; therefore, all we need to do is update $t$.
In the remaining case, the small bin containing $C$ in $O'$ may lead to a \emph{bad} performance, 
so we make some changes to the arrival times or deadlines of some items such that, after the modification, the bin containing $C$ is no longer small.
The details of how to perform the modification in each case will be discussed in~\cref{sec:details_special_case}. 

We will show that, after each update, the following invariants hold:

\begin{itemize}
  \item Invariant 1:  $I$ and $I'$ use the same set of items, except some items may have more ``restrictive'' arrival times and deadlines in $I'$.
  \item Invariant 2:  For bins with packing time after $t$, $x' - z' \leq x^*$.
  \item Invariant 3:  $\ALG(I) \leq \ALG(I')$.
  \item Invariant 4:  Time $t$ will be moved to an earlier time slot.
\end{itemize}

The first three invariants correspond to the three properties mentioned in \cref{sec:special_case}, 
except we only require that the original second property holds only for bins packed after~$t$.  
Invariant 4 guarantees that $t$ will be moved to an earlier time slot after each update, 
so the conversion algorithm will terminate.

It is easy to verify that all the invariants hold initially, when $t$ is set as the latest deadline of all items.   
Also, when the conversion terminates, there are no small bins with packing time \emph{before} $t$, and so the number of small bins $x'$ will not change even if we move $t$ to the earliest moment.
That is, Invariant 2 will continue to hold not just for bins after $t$, but also for all bins.
In other words, when the conversion terminates, the sequence $I'$ obtained will satisfy all three properties in \cref{sec:special_case}. 

It remains to describe how each case is handled and to show that the four invariants are maintained after the update.

\medskip

\begin{algorithm}
\caption{Input Conversion in Special Case}
\label{alg:special_case}
\begin{algorithmic}[1]
\renewcommand{\algorithmicrequire}{\textbf{Input:}}
\renewcommand{\algorithmicensure}{\textbf{Output:}}
\REQUIRE Sequence $I$
\ENSURE  Modified sequence $I'$

\STATE Fix any optimal algorithm $\OPT$;
\STATE $O^* \leftarrow$ output by $\OPT$ on $I$;
\STATE $I' \leftarrow I$;
\STATE $O' \leftarrow$ output by $\ALG$ on $I'$;
\STATE $t \leftarrow$ latest deadline of all items in $I'$;
\WHILE {$O'$ contains a small bin with packing time before $t$}
    \STATE $b \leftarrow$ the small bin in $O'$ with latest packing time before $t$;
    \STATE $C \leftarrow$ trigger item of $b$;
    \STATE Update $t$ and $I'$ based on $C$, $O'$ and $O^*$; \COMMENT{Details in \cref{sec:details_special_case}}
    \STATE $O' \leftarrow$ output by $\ALG$ on $I'$;
\ENDWHILE
\RETURN $I'$ 
\end{algorithmic} 
\end{algorithm}

\subsection{Details of Updates (Line 9)}
\label{sec:details_special_case}

\subsubsection{Case 1}
Here, the item $C$ is also in a small bin in $O^*$.
When this happens, we do nothing for the items in $I'$,
and just move $t$ to the arrival time of $C$.

As $I'$ is not changed, Invariants $1$ and $3$ will hold automatically
after the update.  Also, $t$ is moved to an earlier moment, so Invariant 4 also holds.
To see why Invariant 2 also holds, let us fix some notation.
Let $t_0$ denote the value of $t$ before the update, and $t_1$ denote the arrival time of $C$, which is the value of $t$
after the update.
Let $x'_0$ and $x'_1$ denote the number of small bins packed after $t_0$ and after $t_1$ in $O'$, respectively.  
Similarly, let $z'_0$ and $z'_1$ denote the number of large bins packed after $t_0$ and $t_1$ in $O'$,
and $x^*_0$ and $x^*_1$ denote the number of small bins packed after $t_0$ and $t_1$ in $O^*$.

By the choice of $C$, and the fact (\cref{lem:trigger}) that the intervals of trigger items of small bins in $O'$
do not overlap, we conclude that $x'_1 = x'_0 + 1$.  As $t_1$ is an earlier moment than $t_0$,  $z'_0 \leq z'_1$.
Also, $C$ is in a small bin in $O^*$, so we further have $x^*_1 \geq x^*_0 + 1$.  Now, because Invariant 2 holds after $t_0$,
we have $x'_0 - z'_0 \leq x^*_0$.  Immediately, this implies 
\[  x'_1 - z'_1\ \leq\ x^*_1, \]
so that Invariant 2 also holds after $t_1$.

\subsubsection{Case 2}
Here, some other bin is packed in the interval of $C$ in $O'$.
When this happens, we do nothing for the items in $I'$,
and just move $t$ to the arrival time of~$C$.

Similarly to Case 1, we see that Invariants 1, 3 and 4 will hold after the update.
To see why Invariant 2 also holds, let us re-use the notation introduced in Case 1.
That is, 
\begin{itemize}
  \item $t_0$ and $t_1$: the value of $t$ before and after the update;
  \item $x'_0$ and $x'_1$: the number of small bins packed after $t_0$ and after $t_1$ in $O'$;
  \item $z'_0$ and $z'_1$: the number of large bins packed after $t_0$ and $t_1$ in $O'$; and
  \item $x^*_0$ and $x^*_1$: the number of small bins packed after $t_0$ and $t_1$ in $O^*$.
\end{itemize}

Similar to Case 1, we have $x'_1 = x'_0 + 1$.  Next, as there is some other bin $b$ packed
in the interval of $C$, such a bin cannot be a small bin nor a medium bin, by \cref{lem:trigger}.
Thus, $b$ is a large bin, so that $z'_1 \geq z'_0 + 1$.
As $t_1$ is an earlier moment than $t_0$,  $x^*_0 \leq x^*_1$.
Now, because Invariant 2 holds after $t_0$,
we have $x'_0 - z'_0 \leq x^*_0$.  Immediately, this implies 
\[  x'_1 - z'_1\ \leq\ x^*_1, \]
so that Invariant 2 also holds after $t_1$.

\subsubsection{Case 3}
Here, we consider all those situations that do not match the conditions of Case~1 or Case~2.  
In this case, we will make modifications to the arrival time of a selected item, and possibly the deadlines of some other items.  
At any time, we ensure that the following sub-invariants hold:

\begin{itemize}
   \item \textbf{(SI1)} For any item $p$ whose arrival time is modified, its modified arrival time will be after $t$.
         Furthermore, it is packed with some item $q$ in $O^*$ whose arrival time is after $t$.
   \item \textbf{(SI2)} For any item $p$ whose deadline is modified, its modified deadline will be before its original deadline.
         Furthermore, the modified deadline is on or after $t$.
   \item \textbf{(SI3)} Let $p$ and $q$ be items that are packed before $t$ by our algorithm $\ALG$ on $I'$.
         Then, their time intervals overlap in $I$ if and only if their intervals overlap in $I'$.
\end{itemize}
Obviously, the above sub-invariants are true before any item of $I$ is modified.  In the following, we assume
inductively that they are true before the current modification;  based on that, we describe how the modification is made, 
while ensuring that they continue to hold after the modification, and finally show that the main invariants,
Invariants~1,~2,~3,~4, will also hold after the modification.

Our description now begins.  First, because we are not in Case~1, item $C$ is not in a small bin in $O^*$. This implies that $C$ is packed with some other item, say $B$. 
Note that the arrival times of both $B$ and $C$ have not been modified, as they are not after $t$.
Also, the time intervals of $B$ and $C$ overlap in $I'$, whether their deadlines have been modified 
or not (as a modification on deadline can ``truncate'' only the portion of an interval after $t$).
Then, we have:

\begin{lemma}
The arrival time of $B$ is before the arrival time of $C$.
Also, in $O'$, $B$ is packed in some bin before the arrival time of $C$.
\end{lemma}
\begin{proof}
Consider $O'$. Since we are not in Case 2, no other bin is packed in the interval of $C$. 
This implies that $B$ is either packed before the arrival time of $C$, or after the deadline of $C$.  
Yet, the latter one is impossible, as otherwise $B$ is available by the time $C$ is packed by our algorithm,
so that $B$ would have been put together with $C$ in the same bin.  
In other words, $B$ \emph{must be} packed before the arrival time of $C$ in $O'$.
Furthermore, $B$ must arrive before $C$ for this to occur.
\end{proof}

\begin{figure}[ht]
\centering
\begin{tikzpicture}
\draw[<->] (0,3) -- (2,3) node at (1,3) [above] {$A$};
\draw[-] (1,2) -- (4,2) node at (3.5,2) [above] {$B$};
\draw[<->] (3,1) -- (5,1) node at (3.5,1) [above] {$C:\frac{1}{3}$};
\draw[->, red, very thick] (1,2) -- (3,2);
\end{tikzpicture}
\caption{Intervals of items \(A\), \(B\), and \(C\)
         before and after the update.}
\label{fig:special_case_ABC}
\end{figure}
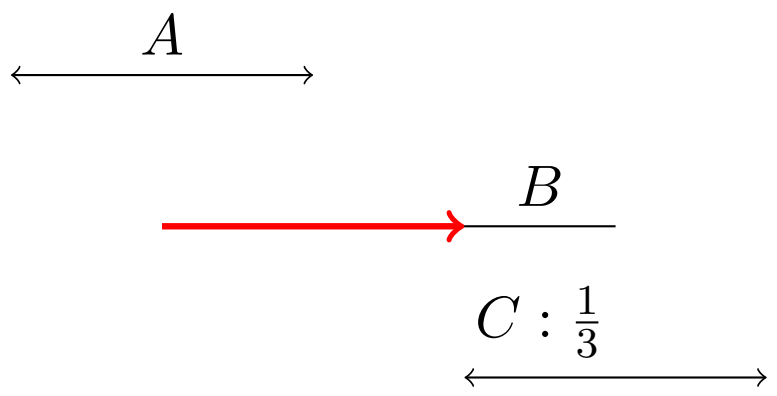

Let us look at the bin $b'$ in $O'$ where $B$ is packed.  
As this bin is packed before the arrival time of $C$, it must be packed before the deadline
of $B$.  So, $B$ is not the trigger item of $b'$.
Instead, there exists some other trigger item,
say $A$. See \cref{fig:special_case_ABC} for the relationship of the intervals of $A$, $B$, and $C$.

Let $[a_A, d_A]$ denote the interval of $A$, and $[a_C, d_C]$ denote the interval of $C$.  
Let $O'_0$ and $O'_1$ denote the output of our algorithm $\ALG$ on $I'$ before and after the update.
Then, the update of $I'$ and $t$ in Case 3 is as follows:
\begin{itemize}
  \item For each item $p$ (except $B$) that is packed before $a_C$ in $O'$,
        if the original deadline is after $a_C$, we set the deadline to be a value just before $a_C$,
        say $a_C^- = a_C - \varepsilon$ for an arbitrarily small positive $\varepsilon$.
  \item We set the arrival time of $B$ to $a_C$.
  \item We set $t$ to $a_C^-$.
\end{itemize}
Briefly speaking, the purpose of the above update is to put $B$ and $C$ together, and also ``freeze'' the bins packed after $t$, so that they are not affected in the future.

It is easy to see that, after the update, all three sub-invariants (SI1, SI2, and SI3) continue to hold.
As for the main invariants, trivially, Invariants 1 and 4 will hold after the update.  
For Invariants 2 and 3, intuitively,
we will show that there is little difference between $O'_0$ and $O'_1$,
where $O'_1$ denotes the output of our algorithm $\ALG$ on $I'$ after the update,
so that the arrangement of items in their bins can be compared directly.  
In particular, we consider the items in different \emph{time zones}, and
we will show that:

\begin{enumerate}
  \item \textbf{(Z1)} 
        items packed within $[0, d_A)$ in $O'_0$
        are exactly those items packed 
        within the same period $[0, d_A)$ in $O'_1$.
        Furthermore, bins are packed in the same way, 
        so that in this period,
        the number of bins packed in~$O'_1$ 
        is equal to the number of bins packed in~$O'_0$.          
  \item \textbf{(Z2)} 
        items packed within $[d_A, a_C)$ in $O'_0$, except $B$,
        are exactly those items packed 
        within the same period $[d_A, a_C)$ in $O'_1$.
        Also, in this period, 
        the number of bins packed in $O'_1$ 
        is at least the number of bins in $O'_0$. 
  \item \textbf{(Z3)} 
        In the period $[a_C, d_C]$, 
        $O'_0$ has exactly one bin, which is packed with $C$ only,
        while $O'_1$ also has exactly one bin, which is
        packed with both $B$ and $C$ (thus, a medium or a large bin).  
  \item \textbf{(Z4)} 
        items packed after $d_C$ in $O'_0$
        are exactly those items packed 
        after $d_C$ in $O'_1$.
        Furthermore, bins are packed in the same way, 
        so that in this period,
        the number of bins packed in $O'_1$ 
        is equal to the number of bins packed in $O'_0$.
\end{enumerate}

The above statements imply that after the update, the number of bins in $O'_1$
is at least the number of bins in $O'_0$, so that Invariant 3 will hold after the update.
Furthermore, let:
\begin{itemize}
  \item $t_0$ and $t_1$: the value of $t$ before and after the update;
  \item $x'_0$ and $x'_1$: the number of small bins packed after $t_0$ in $O'_0$ 
                           and after $t_1$ in $O'_1$;
  \item $z'_0$ and $z'_1$: the number of large bins packed after $t_0$ in $O'_0$ 
                           and after $t_1$ in $O'_1$; and
  \item $x^*_0$ and $x^*_1$: the number of small bins packed after $t_0$ and $t_1$ in $O^*$.
\end{itemize}
Then, statements Z3 and Z4 imply that 
\[   x_1'\ =\ x_0' \quad\text{and}\quad  z_0'\ \leq\ z_1'. \]
This is because the bin containing $C$ will no longer be small, but may become large after the update,
while all the other bins packed after $t_1 = a_C$ are the same in both outputs $O'_0$
and~$O'_1$.  
Also, as $t_1$ is moved to an earlier moment,
\[  x^*_0\ \leq\ x^*_1. \]
As Invariant 2 holds after $t_0$ in $O'_0$, we have $x'_0 - z'_0 \leq x^*_0$.  
Immediately, this implies 
\[  x'_1 - z'_1\ \leq\ x^*_1, \] so that Invariant 2 also holds after the update.

To complete the whole discussion, it remains to show that the above statements for the different
time zones are true.  Details are as follows.

\paragraph{Proof of Z1:} 
Since the input restricted to the interval $[0,d_A)$ remains unchanged after the update, 
and by sub-invariant SI3 the overlap relations among these items are preserved, 
re-running $\ALG$ produces exactly the same bins with packing times in $[0,d_A)$.

\paragraph{Proof of Z2:} We first observe the following result.

\begin{lemma}\label{lem:Z2-trigger}
    For bins packed within the interval $[d_A,a_C)$ in $O'_0$, every trigger item remains a trigger item in $[d_A, a_C)$ in $O'_1$.
\end{lemma}
\begin{proof}
    Let $I'_0$ and $I'_1$ denote the sequence $I'$ before and after the update, respectively.
    Starting from time $d_A$ where we open a bin containing $A$,
    since $a_B$ is modified, the bin will no longer contain $B$, and the items inside may become different.
    Yet, any items added to the bin before $B$ in $O'_0$ will still be in the bin in $O'_1$.
    Thus, only an item, say $p$, that is added after $B$ in $O'_0$ may not be added to the bin now,
    because another item $q$ is added to the bin instead.  Yet, both $p$ and $q$ must have deadline after that of $B$
    in $I'_0$, for otherwise they will be added to the bin before $B$ in $O'_0$.
    As the deadline of $B$ in $I'_0$ is after $a_C$, the deadline of $q$ is also after $a_C$,
    so $q$ cannot be a trigger item in $[d_A, a_C)$.
    Also, for this to happen, the deadline of $p$ must be after $q$.

    Here, we call $q$ the ``culprit'' and $p$ a ``victim''.  Note that there can be more than one victim.

    As time goes on, we will open a bin for the next item whose deadline is due.
    The other items, if they exist, in the bin would be the same as in $O'_0$ in general, 
    but when some of them, say $q$, becomes a culprit of an earlier bin, 
    the items inside may become different.
    Yet, any items added to the bin before $q$ in $O'_0$ will still be in the bin in $O'_1$
    if they are available now.
    Thus, only an item, say $p'$, that is added after $q$ in $O'_0$ may not be added to the bin now, 
    because another item $q'$ is added to the bin instead.  With the same argument as before,
    the new culprit $q'$ cannot be a trigger item in $[d_A, a_C)$, $p'$ has deadline after $q'$,
    and all culprits and victims in the process so far have deadlines after that of $B$ in $I'_0$.

    As time goes on, more and more culprits and victims may be formed.
    Yet, \emph{none} of them can be a trigger item of any bin in $O'_0$.
    Also, as they have deadlines after that of $B$ in $I'_0$, 
    those deadlines will be after $a_C$ in $I'_0$;
    consequently, they cannot be a trigger item in a bin within $[d_A, a_C)$ in $O'_0$.
    In other words, only non-trigger items can be culprits or victims,
    so that they are the only items to be ``re-distributed'' into a different bin,
    while all trigger items in $O'_0$ in $[d_A, a_C)$ must still remain in their own bins,
    and be trigger items in $O'_1$.
\end{proof}

By~\cref{lem:Z2-trigger}, the number of bins in [$d_A, a_C$) in $O'_1$ is at least the number of bins in $O'_0$.
Also, for all items packed during this period in $O'_0$, their deadlines are now before $a_C$ after the update,
so that they will all be packed during this period in $O'_1$.

\paragraph{Proof of Z3:}
By statement Z1 and statement Z2, all items with packing time before $a_C$ (except B) are packed before $a_C$ in $O'_1$, 
$B$ and $C$ will be the only items left unpacked during $[a_C, d_C]$, 
so that there will be exactly one bin in $O'_1$ in this period, containing them.

\paragraph{Proof of Z4:}
By statements Z1, Z2, and Z3, all items with packing time before $d_C$ are still packed before $d_C$, 
so for the time after $d_C$, all unpacked items are the same in $I'_0$ and in $I'_1$.
Also, neither their arrival times nor their deadlines are updated.
Thus, after re-running $\ALG$, all the bins with packing time after $d_C$ will be exactly the same.
\section{Conclusion}
\label{sec:conclusion}

In this paper, we revisited online \TimelyBinPacking{}.
We proposed a simple deterministic algorithm with a competitive ratio of exactly $2$, matching the best-known upper bound, while simultaneously achieving the optimal competitive ratio for two classes of inputs.

Several interesting problems remain open:
\begin{itemize}
\setlength{\itemsep}{-2pt}
  \item The optimal competitive ratio for deterministic online algorithms remains between $1.5$ and~$2$. 
  Closing this gap appears challenging.

  \item This paper focused on deterministic algorithms, but randomization may lead to better performance.
  Aminian~\cite[Theorem~4.2.2]{aminian2025algorithms} proved that, for inputs of length $n$, no randomized online algorithm can achieve a competitive ratio below $1.5-\frac{1}{2n-4}$.
  This lower bound approaches $1.5$ as $n$ grows.
  It remains open whether a randomized online algorithm can achieve a competitive ratio strictly below $2$.

  \item In \TimelyBinPacking{}, pending items may be stored temporarily without cost and without any limit on storage capacity.
  It would be interesting to study variants with bounded storage or with holding costs for pending items.
\end{itemize}
\section*{AI Methodology}

ChatGPT was used to assist with literature searches and with improving the authors’ existing text.

\printbibliography

\appendix
\crefalias{section}{appendix}
\crefalias{subsection}{appendix}

\section{Lower Bounds for Discrete-Greedy}
\label{appendix:Discrete_Greedy_LB}
The $\DiscreteGreedy$ algorithm works by simulating the optimal fractional algorithm, $\Greedy$, such that each bin packed by $\Greedy$ is projected to no more than two integral bins.

For example, suppose $\Greedy$ packs the bins $b_1, b_2, b_3, b_4$, where the last item $p$ of $b_1$ is split between $b_1$ and $b_3$, and the last item $q$ of $b_3$ is split between $b_3$ and $b_4$. Then, $\DiscreteGreedy$ produces: $b_1$ as before but without the fraction of $p$; a dedicated bin $b'_1$ for $p$; $b_2$ as before; $b_3$ as before but without the fractions of $p$ and $q$; a dedicated bin $b'_3$ for $q$; and $b_4$ as before but without the fraction of $q$.

The projection step makes the analysis for the competitive ratio of $\DiscreteGreedy$ extremely simple. However, the projection step also leads to inefficiency. As we now show, $\DiscreteGreedy$ has a competitive ratio of at least $2$, even for two simple classes of inputs.

\subsection{Uniform-Size Items}
We construct a sequence of inputs $I_n$ such that all items in $I_n$ have the same size (for each $n$) and ${\DiscreteGreedy(I_n)} \to 2 \, {\OPT(I_n)}$ as $n \to \infty$, showing that $\DiscreteGreedy$'s competitive ratio is at least $2$ for this class of inputs.

Let $\varepsilon$ be a real number such that $0 < \varepsilon \le \frac{1}{n(n+1)(n+2)}$. The input $I_n$ consists of items with arrival time $0$ and size $s = \frac{1}{n+1} + \varepsilon$. The items are divided into $n+1$ groups:
\begin{itemize}
    \item Groups $1, \dots, n$: Each group contains $n+1$ items. Group $i$ has deadline $i$.
    \item Group $n+1$: Contains $n$ items with deadline $n+1$.
\end{itemize}

For $\Greedy$, items are sorted by increasing deadlines. Since $(n+1)s > 1$, a single bin cannot hold all items of one group. Hence, $\Greedy$ fills the first $n$ bins $b_1, \dots, b_n$ completely. Due to the small $\varepsilon$, the total size of all items is no more than $n+1$, so $\Greedy$ uses $n+1$ bins. Notice that since the cumulative size of the first $k$ groups exceeds $k$ by a small margin $k(n+1)\varepsilon < s$, each bin $b_1, \dots, b_n$ ends with a split item.

Consequently, $\DiscreteGreedy$ projects each of the bins $b_1, \dots, b_n$ into 2 integral bins (one for the non-split items, one for the split item that closes the bin) and projects $b_{n+1}$ into 1 integral bin. Thus, $\DiscreteGreedy$ uses $2n + 1$ bins.

For $\OPT$, each bin holds exactly $n$ items (since $ns < 1 < (n+1)s$). The total number of items is $n(n+1) + n = n(n+2)$. Thus, $\OPT$ can pack all items into $\lceil \frac{n(n+2)}{n} \rceil = n + 2$ bins.

\subsection{Items of Sizes in \texorpdfstring{$\{1/3,2/3\}$}{\{1/3, 2/3\}}}
Similarly, we construct a sequence of inputs $I_n$ such that all items have size either $1/3$ or $2/3$, and ${\DiscreteGreedy(I_n)} \to 2 \, {\OPT(I_n)}$ as $n \to \infty$, showing that $\DiscreteGreedy$'s competitive ratio is at least $2$ for this class of inputs.

The input $I_n$ consists of items with arrival time $0$. The items are divided into $n+1$ groups, ordered by increasing deadlines as follows:
\begin{itemize}
    \item Group $1$: Two items of size $2/3$.
    \item Groups $2, \dots, n$: Each group contains one item of size $1/3$ followed by one item of size $2/3$.
    \item Group $n+1$: One item of size $1/3$.
\end{itemize}

For $\Greedy$, the items are packed in the order specified above. 
\begin{itemize}
    \item The first bin $b_1$ packs the first item of Group 1 (size $2/3$) and the first $1/3$ of the second item. It ends with a split item.
    \item For $k = 2, \dots, n$, the bin $b_k$ starts with the remaining $1/3$ of the split item from Group $k-1$, packs the item of size $1/3$ from Group $k$, and fills the remaining space with the first $1/3$ of the item of size $2/3$ from Group $k$. Thus, each bin $b_k$ is full and ends with a split item.
    \item The final bin $b_{n+1}$ starts with the remaining $1/3$ of the split item from Group $n$ and packs the single item of size $1/3$ from Group $n+1$. The bin is not full but contains no split items at the end.
\end{itemize}
In total, $\Greedy$ uses $n+1$ bins. Since bins $b_1, \dots, b_n$ all end with a split item, $\DiscreteGreedy$ projects each of them into 2 integral bins (one for the non-split item, one for the split item that closes the bin). The final bin $b_{n+1}$ projects to 1 integral bin. Thus, $\DiscreteGreedy$ uses $2n + 1$ bins.

For $\OPT$, we can pair the items. There are $n+1$ items of size $2/3$ and $n$ items of size $1/3$. Thus, $\OPT$ uses $n+1$ bins.

\end{document}